\documentclass{emulateapj-rtx4}
\usepackage{graphicx}

\slugcomment{Manuscript Version Date: \today}

\shorttitle{High Resolution NIR and MIR Imaging of Maser Rings}
\shortauthors{De Buizer, Bartkiewicz, \& Szymczak}

\begin{document}

\title{Testing the Hypothesis that Methanol Maser Rings Trace Circumstellar Disks:\\ High Resolution Near-IR and Mid-IR Imaging}

\author{James M. De Buizer}
\affil{Stratospheric Observatory For Infrared Astronomy -- USRA\\
 NASA Ames Research Center, MS N232-12, Moffett Field, CA 94035, USA}
\email{jdebuizer@sofia.usra.edu}
\and
\author{Anna Bartkiewicz and Marian Szymczak}
\affil{Toru\'n Centre for Astronomy, Nicolaus Copernicus University, Gagarina 11, 87-100 Toru\'n, Poland}

\begin{abstract}
Milliarcsecond VLBI maps of regions containing 6.7~GHz methanol maser emission have lead to the recent discovery of ring-like distributions of maser spots and the plausible hypothesis that they may be tracing circumstellar disks around forming high mass stars. We aimed to test this hypothesis by imaging these regions in the near and mid-infrared at high spatial resolution and compare the observed emission to the expected infrared morphologies as inferred from the geometries of the maser rings. In the near infrared we used the Gemini North adaptive optics system of Altair/NIRI, while in the mid-infrared we used the combination of the Gemini South instrument T-ReCS and super-resolution techniques. Resultant images had a resolution of $\sim$150\,mas in both the near-infrared and mid-infrared. We discuss the expected distribution of circumstellar material around young and massive accreting (proto)stars and what infrared emission geometries would be expected for the different maser ring orientations under the assumption that the masers are coming from within circumstellar disks. Based upon the observed infrared emission geometries for the four targets in our sample and the results of SED modeling of the massive young stellar objects associated with the maser rings, we do not find compelling evidence in support of the hypothesis that methanol masers rings reside in circumstellar disks.
\end{abstract}


\section{Introduction}

The formation of OB stars is an important challenge to modern astrophysics as they are responsible for many of the energetic phenomena in galaxies. But their large distances, heavy obscuration and rapidity of evolution make studies of massive star formation difficult.
One property that is unique to massive young stellar objects is the presence of methanol masers in their early formative phases \citep{m91,c95}.  As such, they may hold clues to understanding the fundamental differences between low and high mass star formation.  High angular resolution observations of the masers allow us to map the neutral gas at scales of a few 10 AUs in the vicinity of massive young stars.  However, because of their relatively sparse sampling, it is still unclear where and how the masers are formed.  Two competing hypotheses have arisen; one stating that methanol masers are embedded in circumstellar tori or accretion disks around the massive protostars (e.g., Norris et al. 1993), the other stating that they may generally be tracing outflows \citep{d03}.

However, milliarcsecond scale very long baseline interferometry (VLBI) observations show a wide range of morphologies for the maser spot distributions at 6.7\,GHz.  They form simple structures (a single maser spot), or group into spots that lie in lines or arcs, or are distributed randomly without any regularity \citep{m00,n98,p98,w98}.  Linear structures of maser emission accompanied by clear velocity gradients have led to the belief that they may be in edge-on disks \citep{m00,w98}.  Other scenarios where outflows or shocks are generating the linear or arc-like distributions have also been proposed \citep{d04}.

We have recently completed a survey of 31 sources at 6.7\,GHz using European VLBI Network (EVN) \citep{b09}.  In addition to the curved and complex morphologies observed in other samples, we have discovered for the first time nine sources (29\,\% of the sample) with ring-like maser distributions with typical sizes having major axes of 0\farcs2-0\farcs3.  Though not apparent in the radio data, these ring-like structures strongly suggest the existence of a central stellar object, and prompt the obvious question: {\it Are these maser rings tracing disks around massive protostars?}

Inspection of mid-infrared (MIR) data from the {\it Spitzer} IRAC  maps, GLIMPSE and MIPSGAL\footnote{http://irsa.ipac.caltech.edu/}, revealed that all of our sources with ring-like morphologies coincide with unresolved MIR sources within one pixel in a GLIMPSE map (1\farcs2).  Many sources also have bright near-infrared (NIR) counterparts unresolved in the 2MASS survey (Cutri et al. 2001).  However, the resolutions of the {\it Spitzer} data (2\farcs3 at 8\,$\mu$m, 7$\arcsec$ at 24\,$\mu$m) and 2MASS data (2.5\arcsec) are inadequate in helping to understand the detailed environment of these maser rings.  Furthermore, several of the sources as seen by {\it Spitzer} are saturated and are so bright that they cause image artifacts preventing one from knowing the true nature of any extended dust emission, if any. Therefore, we decided to obtain the highest spatial resolution NIR and MIR imaging available to explore the true nature of the maser rings.  Using the Gemini 8-m telescopes, we obtained data with NIRI/Altair, an adaptive optics NIR instrument that can achieve resolutions of $\lesssim$150 mas at 2~$\mu$m. MIR observations were made with T-ReCS, employing a method which fully characterized the system point spread function accurately enough to allow the imaging data to be reliably deconvolved, achieving spatial resolutions of $\sim$150 mas at 8~$\mu$m and $\sim$250 mas at 18~$\mu$m.

To better understand the nature of the methanol rings, the main aim of our observations was to resolve the circumstellar dust emission at NIR and MIR wavelengths to verify whether or not their infrared morphologies are consistent with the hypothesis that methanol maser rings are tracing circumstellar disks. Additionally, through the use of spectral energy distribution (SED) model fitting, we will learn more about the possible physical properties and source geometries of the massive young stellar objects (MYSOs) associated with the maser rings.

\section{Observations}
\subsection{Sample selection}
The targets were selected from the sample presented in \citet{b09}. Nine out of 31 methanol masers spot distributions in that study can be fit by ellipses, but of those, only seven targets have four or more maser clumps and thus are the most convincing maser rings in the survey. Inspection of archived {\it Spitzer} data revealed that there are MIR counterparts to all seven maser rings.  However, in four out of these seven targets the MIR emission in the {\it Spitzer} maps at 8 and 24\,$\mu$m is saturated, blowing out the centers of the sources and creating ring-like ghost structures.  This makes it impossible to use the {\it Spitzer} data to derive any information about the
spatial extent and/or shape of the MIR emission, however this did indicate that the target would certainly be bright enough to be observed with reasonable exposure times on less sensitive ground-based telescopes.  The four maser ring target fields are given in Table 1.

\subsection{MIR observations}
We used the Thermal-Region Camera and Spectrograph (T-ReCS) at Gemini South in queue mode (Program ID GS-2009B-Q-7), with data being obtained on several nights spanning the 2009 July 29 to 2009 August 29 time period.  Observations were taken with the $Si$-2 ($\lambda_c$=8.74~$\mu$m, $\Delta\lambda$=0.78~$\mu$m) and $Qa$ ($\lambda_c$=18.30~$\mu$m, $\Delta\lambda$=1.6~$\mu$m) filters.  T-ReCS employs a Raytheon 320$\times$240 pixel Si:As IBC array which is optimized for use in the 7--26~$\mu$m wavelength range.  The pixel scale is 0.089$\arcsec$/pixel, creating a field of view of 28$\farcs$8$\times$21$\farcs$6.  Total exposure times were 340\,s for the 8.6~$\mu$m filter and 360\,s for the 18.3~$\mu$m filter.  Standard chop-nod observing was used to remove sky and telescope radiative offsets.

Flux calibration was achieved by observing the mid-infrared standard star HD168723 ($\eta$ Ser) at a similar airmass as each of the science targets. The assumed flux densities were taken to be 20.67 Jy at 8.6 $\mu$m and 4.98 Jy at 18.3 $\mu$m.  These assumed flux densities were found by convolving the T-ReCS filter profiles with the spectral irradiance templates of Cohen et al.  (1999).  Variability in the measured response of the standard star from night to night was used to estimate the error in the flux density observations.  From this we conclude that the measured 8.6 $\micron$ flux densities have a 1-$\sigma$ error of 8\% and the 18.3 $\mu$m flux densities have an error of 15\% (see Table 2).

\subsubsection{MIR Deconvolution and Resolution}
Our observations at 8.6 and 18.3~$\mu$m yielded natural spatial resolutions of $\sim$0.29$\arcsec$ and $\sim$0.50$\arcsec$, respectively. Since the delivered point-spread function (PSF) can change due to different wind loads on the secondary mirror, observations of bright mid-infrared stars were taken immediately before and/or after each science target observation. Furthermore the PSF can change due to differences in the telescope flexure as a function of telescope position, so the PSF stars were chosen to be no further than 1$\degr$ away from the science targets. From these observations we believe that we can accurately characterize the PSF of the science observations, and using these PSFs, deconvolve the target images using the maximum likelihood method \citep{r72,l74}. That increases the effective resolution of our images by a factor of $\sim$2 (i.e., $\sim$0.15$\arcsec$ at 8.6~$\mu$m and $\sim$0.25$\arcsec$ at 18.3~$\mu$m).  G23.657-00.127 and G23.389+00.185 yielded detections of unresolved point sources in the MIR, and thus their images were not deconvolved. However, G24.634-00.324 and G25.411+00.105 displayed extended MIR emission and therefore we deconvolved both the 8.6 and 18.3\,$\mu$m images of these sources. These deconvolved images compare favorably to simple unsharp masking of the original images, and hence all of the substructures revealed in the deconvolved images are believed to be real with high confidence. In the case of G25.411+00.105, the deconvolution was also performed on the 2.12~$\mu$m image to enhance the view of the complex structure observed there.

\subsection{NIR observations}
Three of the four maser rings (G23.389+00.185, G23.657-00.127, and
G25.411+00.105) had bright enough 2MASS $K$ band counterparts ($m_K < 12.9$) to be observed with Gemini in reasonable exposure times.  In order to image the sources in the highest spatial resolution possible at these wavelengths we employed the use of near-infrared adaptive-optics (AO) imaging using the ALTtitude conjugate Adaptive optics for the InfraRed (ALTAIR) in conjunction with the Near InfraRed Imager and Spectrometer (NIRI) at Gemini North. Observations were carried out from 8 August to 16 October 2009 in queue mode (Program ID GN-2009B-Q-44) using the $K^{\prime}$ ($\lambda_c$=2.12~$\mu$m, $\Delta\lambda$=0.35~$\mu$m) filter.  NIRI has a 1024$\times$1024 ALADDIN InSb array and has three selectable pixel scales and fields of view.  We selected the f/32 mode which yields a pixel scale of 0.0219$\arcsec$/pixel and a field of view of 22.4$\arcsec$$\times$22.4$\arcsec$.  Because all three sources lie in generally extinguished regions, there were no stars bright enough to guide with nearby, and the laser guide star (LGS) system was used.  The LGS system does not allow the AO system to make corrections as well as it could with a natural guide star.  Furthermore, the nights of our observations had less than optimal seeing.  Though the NIRI/Altair system is capable of a top resolution of 0.07$\arcsec$ at 2~$\micron$, our observations have resolutions in the 0.11$\arcsec$--0.16$\arcsec$ range.

The $K^{\prime}$ band AO imaging was performed using a 3 by 3 dither pattern with 4$\arcsec$ offsets.  The median of these nine images was used to produce a sky frame that was subtracted from each of the frames before constructing the final co-added mosaic image.  The effective final exposure times for the three sources is 756\,s for G23.389+00.185, 900s for G23.657-00.127, and 1080\,s for G25.411+00.105.  Flux calibration was achieved by observing a nearby standard star either just before or just after each science observation.  The standard star GSPC S875-C ($m_{K^{\prime}}=10.760$) was used for G23.657-00.127 and G25.411+00.105, and FS 148 ($m_{K^{\prime}}=9.441$) was used for G23.389+00.185.  The measured flux densities of the calibrators varied by only 1\% from night to night, and the derived $K^{\prime}$ band flux densities for several sources on each field were compared to their $K$ values as given in the 2MASS Point Source Catalog.  From this we estimate the absolute calibration errors in the measured $K^{\prime}$ band flux densities of our science targets to be better than 5\% (See Table 2).

\subsection{NIR and MIR Astrometric Accuracies}

Gemini observatory has developed a technique that allows their optical telescopes to be pointed accurately at objects with no optical emission. The technique involves the use of two optical astrometric standard stars, one 5--7$\arcmin$ away from and the other $<$50$\arcsec$ away from the non-optical science target.  The location of the non-optical science target is in essence ``triangulated'' using these stars.  The astrometric standard stars were chosen from the Second U.S.  Naval Observatory CCD Astrograph Catalog (UCAC2) catalog, which has positional errors of about 20 mas for stars with $m_R \sim 10-14$. Repeated trials using the set ups with optically bright science targets showed that the 1$\sigma$ deviation was 0.06$\arcsec$.

Since it was of the highest importance that we know precisely where the maser rings are with respect to the dust emission in the MIR, we employed this astrometric technique to our observations.  This was only done for the science observations at 8.6\,$\mu$m.  The 18\,$\mu$m astrometry was achieved by registering the source positions at 18\,$\mu$m with the 8.6\,$\mu$m emission.  If the emission peak at 8.6\,$\mu$m is actually coincident with the 18\,$\mu$m peak, we estimate our absolute positional uncertainty is slightly worse at 18.3\,$\mu$m ($\sim$0.10$\arcsec$) than 8.6\,$\mu$m.

For the three sources that were also observed in the NIR, we employed a different astrometric technique. Since the NIRI fields containing the science targets also contained multiple NIR stars from the 2MASS Point Source Catalog, we were able to use their measured positions to accurately define absolute astrometry of the NIRI images.  A $\chi^2$-minimization technique was used to register the 2MASS fields to the NIRI fields.  Then the offsets in RA and Dec between the final positions of the centroids of the NIRI sources with respect to the 2MASS sources were measured.  The standard deviation in these right ascension and declination offsets yields the 1$\sigma$ error in the absolute astrometry of the NIR images (see Table 1).

Since we are comparing the offsets of infrared emission from the methanol masers, we point out that the error in the absolute coordinates for the methanol masers is on the order of a few mas owing to the phase-referencing technique that was employed (Bartkiewicz et al. 2009). This means that when comparing the maser and infrared positions, the infrared astrometric errors quoted here for each source dominate in all cases.

\section{Discussion and Results}

In order to fully understand how our NIR and MIR imaging can test the hypothesis of whether or not methanol maser rings trace circumstellar disks around massive young stellar objects, we must first state what we expect the IR emission from such a source to look like under the assumption that it is forming via disk/envelope accretion. We will discuss here how the observed properties of embedded young massive stars with disks differ from more well-known IR observations of disks around low mass stars. We will then discuss the expected IR morphologies of a MYSO, and how that morphology differs with angle-dependent disk geometries. We will then discuss the results of our high resolution IR observations and how they compare to the expected IR morphologies of the four MYSOs whose disk geometries were derived from the properties of their methanol maser rings. Finally, we will use SED models to fit the multi-wavelength photometry of each source to derive the likely physical and geometrical properties of each MYSO and compare those results to the expected properties based on the masers and to the observed properties of the IR morphologies seen in the images.

\subsection{The manifestation of IR emission from a disk-accreting massive young stellar object}

The first infrared observations to resolve disks around stars were of debris disks (e.g., Telesco et al. 1988, Jayawardhana et al. 1998). Debris disks are circumstellar disks that occur around stars with more evolved, (post)planet building disks. Since one can directly view a debris disk in its infrared dust emission it appears as an elongated ellipsoidal structures whose ``flatness'' is a function of the viewing angle of the disk. However, at earlier stages of star formation, such circumstellar disks are likely to be actively accreting on to their parent stars, and are thus much more massive and dense. In fact, accretion disks around very young stars began to be found, not by their direct emission, but by their scattered and reflected emission off of their upper and lower disk surfaces. Originally discovered in the optical (e.g., McCaughrean \& O'Dell 1996), these ``silhouette'' disks demonstrated that if a disk is large and dense enough, it could be optically thick even at infrared wavelengths (e.g., McCaughrean et al. 1998, Cotera et al. 2001). These sources are typified by a ``dark lane'' demarcating the disk itself, between two ``lobes'' of scattered/reradiated emission demarcating the upper and lower flared disk surfaces.

However, all of these observations were of disks around low-mass stars. The first claims of infrared detections of circumstellar disks around young massive stars (Stecklum \& Kaufl 1998, De Buizer et al. 2000) were made, in part, on the assumption that the disk would appear as an elongated structure in its infrared dust emission, similar to what was seen with debris disks around low-mass stars. However, higher resolution follow-up observations (De Buizer et al. 2002) proved that these were not disks, and it was soon after realized that disks accreting onto young massive (proto)stars would be optically thick in the infrared as well, and would be even more dense and massive than those accreting onto low-mass protostars. Furthermore,  in order to have a large enough reservoir of material to accrete to such high masses, massive stars must form in the densest parts of giant molecular clouds, which are extremely obscured environments. Moreover, the earliest stages of massive star formation are deeply self-embedded; the (proto)stars are surrounded by massive and dense accretion envelopes, as well as the aforementioned circumstellar disks, which are believed to be large, thick, and flared.

Due to the high amount of obscuration, one would have no hope of observing these stages of massive star formation at even MIR wavelengths if not for their bipolar outflows. Disk accretion is accompanied by outflow, and it is this outflow that punctures holes through the obscuring material surrounding a forming massive star. The outflow axis is oriented perpendicular to the disk, and the outflow starts out narrow and collimated. At such early stages of formation, we may only detect the presence of the massive young stellar object at infrared wavelengths if we are lucky enough to have a line of sight looking down the outflow axis. We could then see into the envelope, and view the scattered/reprocessed dust emission off of the cavity walls, and if the angle is just right, we may see down into the central disk or (proto)star. Such a chance alignment is rare, and it is in part due to this geometrical effect that detecting massive young stars in the infrared is so difficult. However, over time the young stellar object evolves and the outflow angle widens (Shu \& Adams 1987, Beuther \& Shepherd 2005). With the widened outflow, comes a large range of angles that allow for a higher probability of detection in the infrared. At some point the angle is so wide, that the distinction between what is the outflow cavity surface and what is the surface of the flared disk becomes a matter of semantics.

This notional sequence of events has been supported by several studies in the recent decade. For instance, observations of the earliest stages of massive star formation with collimated outflow cavities emitting brightly in their mid-infrared continuum emission were first identified by De Buizer (2006). The first claim of a candidate infrared ``silhouette disk'' around a massive star was made by Chini et al. (2004) demonstrating the later stages where the outflow has widened considerably (though whether the mass of the central object is high mass is the subject of debate, e.g. Sako et al. 2005). In addition to these observations, MYSO radiative transfer models of Alvarez et al. (2004) and Zhang \& Tan (2011) also reproduce these geometries; the first showing the results of the earliest stages of collimated outflow cavities, and the second for more open-angled outflows.

\subsection{Testing the disk hypothesis of maser rings with IR imaging}

Because methanol masers are believed to be pumped by mid-infrared photons from dust (Cragg, Sobolev, \& Godfrey 2002), if the masers are arising from gas in a dusty circumstellar disk, they would have to trace the inner few hundred AU radius at most, given any reasonable heating argument\footnote{Remember however, the flaring of the disk and the dense accretion envelope may prevent one from viewing the mid-infrared emission coming directly from the disk for most, if not all, viewing angles.}. Furthermore the presence of methanol masers is thought to be a signpost of the earliest stages of MYSO formation (e.g., Breen et al. 2010). Additionally, the youngest MYSOs are known to display no cm radio continuum, and none of our sources were found to be associated with 8.4~GHz radio continuum emission (Bartkiewicz et al. 2009).  This means that if the methanol masers in our sample are within circumstellar disks in these sources, then they likely trace a rather young stage of the massive star formation process where the disk is an active accretion disk, likely flared, surrounded by a thick accretion envelope. Figure 1 shows a simplistic toy model of this, where the outflow cavities are the source of the IR emission from the MYSO. The disk and the envelope are assumed to have a uniform opacity (i.e. no formal radiative transfer calculations are implemented), and the cavities are assumed to have uniform emission. Of course these conditions are unrealistic, but show to first order the bulk observable properties that are the result of the more formal modeling of Alvarez et al. (2004) and Zhang \& Tan (2011), and thus are adequate for demonstration purposes.

The toy modeling in Figure 1 shows that there are expected spatial and morphological relationships that can be tested between what one sees in infrared emission (blue and red areas in Figure 1) when looking toward methanol masers rings (green circles) under the assumption that the rings are in disks and that they are associated with an early and embedded accretion phase of massive star formation:
\begin{enumerate}
\item[Scenario 1:] The more circular the maser ring, the more face-on the disk, meaning one would be looking right down the blue-shifted outflow axis into the outflow cavity. Therefore, the infrared emission should be unresolved or circularly symmetric with the peak of infrared emission coincident with the maser ring center;
\item[Scenario 2:] The more highly elliptical the maser ring, the closer to edge-on will be the disk. In the edge-on case the orientation makes it difficult to detect the MYSO in the infrared, unless the envelope is not optically thick to the IR wavelength in question. If detectable in the infrared, we would expect for moderately to highly elliptical maser rings to see something more like a silhouette disk in the infrared, where the maser ring would lie between two infrared bright sources (the outflow cavities), in the ``dark lane'' of the optically thick disk;
\item[Scenario 3:] For an intermediate ellipticity maser ring, one would expect an intermediate disk inclination. In this case we would expect to either only see the blue-shifted outflow cavity, or bright emission from the blue-shifted cavity with a fainter component seen from the red-shifted cavity (with the brightness of the red-shifted cavity dependent on disk/envelope extinction, outflow opening angle, and disk inclination). In this case, the maser ring center will be slightly offset from the infrared emission center in a direction given by the outflow axis.

\end{enumerate}

\subsection{Results from the high resolution IR imaging}

Figure 2 shows the four 6.7\,GHz methanol maser rings in our sample \citep{b09}. Next to each plot of the maser rings the implied disk properties are listed based on fits to the maser rings. Also shown are the toy models created from those same properties, which demonstrate the gross expected infrared emission morphology for each source. In Figures 3--6 we present maps of NIR emission towards three targets (G23.389$+$00.185, G23.657$-$00.127 and G25.411$+$00.105) and MIR emission at 8.6 and 18.3\,$\mu$m towards all four targets. In three out of the four cases we found that the MIR emission at 8.6~$\mu$m is either extended or from multiple components. Furthermore, in all three cases where we also observed the maser rings in the NIR at 2.12\,$\mu$m, we were able to resolve the emission with the high spatial resolution afforded by adaptive optics. In Table 1 we list the coordinates of component peaks detected in the NIR and MIR\footnote{ The naming convention of the IR sources is from the IAU recommendation, which take the form Gll.lll$\pm$bb.bbb:BDS12 \#, where the \# represents the component number. These names are listed in the tables for formality, but in the figures and when discussing components within a particular field are often abbreviated to just the component number alone.}, and in Table 2 we list the integrated flux densities of the components in the field of view at all available bands.

Below we describe results of the infrared observations of each of the four methanol maser rings relevant to the aims of this project. The emphasis here is on the whether or not the NIR and MIR emissions seen are consistent with the scenarios developed in Section 3.2 and toy models shown in Figure 2. There are a few other interesting details that emerged from the data that are not important to the discussion here; these are summarized in the Appendix.

\subsubsection{G23.389$+$00.185}

This source has an inferred disk inclination of 54\degr, so in this case we would expect to see infrared emission distributed as discussed in Scenario 3.

We find a bright unresolved MIR source (source 1) with a peak close (272 mas) to the maser position, and a significantly weaker source (source 2) 2$\arcsec$ ($\sim$9000 AU) to the northeast of this main source in both bands (Figure 3). This second source is far enough away from source 1 that they are likely not two IR-bright cavities from a bipolar outflow of a single MYSO, and are more likely to be separate (proto)stellar sources. Further evidence of this comes from the fact that source 2 appears to have its own jet emanating from it, as seen in the NIR (see Figure 4; details are given in the appendix).

The NIR observations also reveal both sources, with source 1 being a resolved, circularly symmetric source (although this particular NIR data had elongated image quality; see Figure 4). The peaks of source 1 and 2 in the NIR coincide with the peaks in the 8.6~$\mu$m image to better than 60~mas in both cases, and thus the MIR/NIR peaks are believed to be co-spatial.  This demonstrates the effectiveness of both of the two independent astrometric techniques, i.e., the MIR astrometry technique described in the observations section and the ability to triangulate NIR source positions from other 2MASS sources in the field. It also leads to the conclusion that the offset between the maser ring center and infrared peak is significant in this case. In fact, because there was very good spatial agreement between the positions of the 2MASS sources and the NIRI sources on the field in declination, the astrometric technique (described in Section 2.4) led to very small errors along that direction for this source (15\,mas). Since the offset of the maser ring center and the NIR emission center is mostly in declination, that offset is 13$\sigma$.

Moreover, the angle on the sky that the IR peak of source 1 makes with respect to the maser ring center (p.a. = $+$10\degr) is quite different than the disk/outflow axis as inferred from the maser ring geometry (p.a. = $-$45\degr), inconsistent with Scenario 3 and the toy model shown in Figure 2 for this source. However, in Scenario 2 the maser center and IR peak don't have to be coincident, but simply lie along the disk/outflow axis. The closest point along this axis to the IR peak is indeed closer (210 mas), but would still be significantly more than allowed by the combined NIR/MIR astrometry errors ($>$4.5$\sigma$).

The only outflow imaging observations conducted toward this region were the low resolution ($\sim$27$\arcsec$) observations of HCO$+$(3--2) by Schenck et al. (2011). Both IR sources 1 and 2 lie on the southeastern edge of the extended region of integrated HCO$+$ emission, with the peak located $\sim$20$\arcsec$ to the northeast of source 2. Given the coarse resolution it is not clear if the integrated HCO$+$ emission is associated with IR source 1 or 2, and with no channel maps it is unclear if there are well-defined red and blue shifted lobes. It is plausible given the location of the the integrated HCO$+$ peak that it traces the outflow emission associated with the counterjet to the NIR jet seen coming from source 2. If this were the case, the HCO$+$ would be unrelated to the masers associated with IR source 1. If the HCO$+$ emission is tracing an outflow lobe from the YSO associated with the masers, the angle the HCO$+$ peak makes with respect to the maser ring center would imply a outflow direction perpendicular to what is expected from the maser ring geometry.

In summary, the infrared (and HCO$+$) emission from this source does not seem to be consistent with the assumption that the maser ring traces a circumstellar disk.

\subsubsection{G23.657$-$00.127}

For this source, the maser ring is almost circular with an inferred disk inclination of only 16\degr\ from disk model fitting \citep{b05}. It is therefore expected that the IR emission will manifest itself as described in Scenario 1.

Consistent with that scenario, emission from a bright, unresolved source (source 1) is seen in the MIR maps of this maser field (Figure 4). We find that the peak of 8.6\,$\mu$m emission and the peak of the 2.12~$\mu$m emission are coincident with one another to within 0.11$\arcsec$, or 0.8(1.8)$\sigma$ of the NIR(MIR) astrometry error, and thus are likely to be co-spatial. Consistent with Scenario 1 and the toy model for this source, the center of methanol ring is nearly coincident with the NIR(MIR) peak to within 1.0(2.5)$\sigma$.

However, contrary to what is expected in Scenario 1, at 2.12~$\mu$m source 1 is resolved into a extended triangular fan of emission. The top panel of Figure 2 demonstrates that this morphology is real, as there are two other moderately bright sources on the NIRI field that are point-like. While ideally an outflow cavity viewed close to pole-on should have circular symmetry, a non-circular emission morphology is not in itself surprising; in reality, environmental conditions will likely decrease this level of circularity. However for this source, the NIR emission is highly flattened along two sides (see Figure 4), strongly implying the presence of a rather straight barriers of some sort. At present no outflow or molecular imaging of sufficient angular resolution of this object exist to help with the interpretation. However, since the extended fan of emission is seen only in the NIR, the simplest explanation is that it is likely scattered and/or reflected emission off the walls of the outflow cavity. Therefore, given the NIR morphology, this source seems more consistent with a much more inclined disk/outflow than what one would infer from the maser ring geometry.

Therefore, the observed infrared morphologies as a whole are not fully consistent with the hypothesis that the maser ring arises from a face-on circumstellar disk around this source.

\subsubsection{G24.634$-$00.324}

In the case of this maser ring, whose inferred inclination from a fitted disk model is 71\degr, we expect to see an infrared morphology close to that described in Scenario 2.

We do indeed detect a bi-lobed structure in the MIR (Figure 5), with two bright peaks at 18.3 and 8.6\,$\mu$m separated by $\sim$1$\arcsec$ (no NIR data were obtained for this source). This could be interpreted as a silhouette disk-like morphology, where the dark lane between the sources would be the disk midplane. Indeed, as can be seen in the natural resolution 8.6\,$\mu$m image in Figure 5, the methanol maser ring appears to be co-located with the dark lane to within 1.2$\sigma$ of the astrometric uncertainty. The size of the dark lane extends $\sim$1500 AU in diameter, which would be large for a inner accretion disk, but is comparable to what one might expect for a larger-scale outer flared disk or torus (Beuther et al. 2009). The deconvolved 8.6\,$\mu$m image shows the sides of sources 1 and 2 along this dark lane to be very flat, which may be expected if a disk is present, but which would be unusual if these two sources were individual MYSOs. Lending further credence to the silhouette disk-like interpretation is the wavelength dependent nature of the MIR emission. Both sources 1 and 2 grow brighter with wavelength (i.e. 18.3\,$\mu$m compared to 8.6\,$\mu$m), but source 2 increases in brightest more and becomes more extended. This would be expected if source 1 is the side of the disk/outflow facing us and source 2 is the back side partially obscured by the extended and flared disk. The disk becomes less opaque at longer wavelengths, thus would influence the brightening of source 2 more than source 1 as it is observed at longer and longer wavelengths (see Zhang \& Tan 2011). The 1.2 mm continuum maps of this source (Rygl et al. 2010) show a that the cool dust emission in this area peaks $\sim$6$\arcsec$ to the northeast of the maser ring position. However, the mm continuum source is elongated at a position angle of $\sim$60$\degr$, which is approximately the position angle of the MIR ``dark lane'', and extends over $\sim$30$\arcsec$. Optimistically, one could claim that this is consistent with the ``silhouette disk'' interpretation of the IR emission, however because the resolution of the mm maps is only 10.5$\arcsec$, it is not known if the elongation in the mm emission is due to a large torus or simply unresolved multiple dust condensations.

However, if this is indeed a silhouette disk-like morphology, the angle of the outflow/disk axis given by the MIR morphology is 90\degr\ from the axis inferred from the methanol maser ring from the best fitted ellipse. Therefore, the IR morphology and mm dust emission is not consistent with the toy model for this source (as given in Figure 5) and therefore inconsistent with the maser ring tracing a circumstellar disk.

Though we feel that the ``silhouette disk'' disk interpretation is the best explanation for the IR emission given the limited data at hand, it is however plausible that the two sources of IR emission are separate, individual MYSOs. If this were the case then the maser ring lies closer to, and would be more likely associated with, source 2. Some credence to this idea comes from the presence of source 3. This source disappears at longer wavelengths, signifying that the source is hotter and/or less embedded star. This could lead to the idea that all three sources are a tight cluster of YSOs with source 2 being the closest YSO to the masers. However, the maser ring center lies more than 3.2$\sigma$ (of the astrometric uncertainty of the MIR astrometry) away, i.e. much farther away than the ``dark lane''. Ignoring this, if source 2 is an individual MYSO associated with the masers, its emission could be interpreted as the face of a disk/outflow cavity facing towards us with the opposite face/cavity completely obscured from view, in other words the morphology described in Scenario 3. Looking at Figure 5, the MIR peak of source 2 does not appear to lie along the disk/ouflow axis as inferred from the maser ring geometry. However, the maser ring center is only 2.4$\sigma$ away from the closest point along this axis. Thus this scenario could be plausible (however there is even further evidence against this; see Section 3.4), and could be consistent with the maser ring residing in a disk, if the sources of emission seen are individual MYSOs.

Source 3 may also be a foreground or background source, but the 2MASS and Spitzer IRAC images do not have the resolution to confirm this.

On a final note, since this source has only four maser groups fit by an ellipse (see Figure 2), its categorization as a maser ring is the most tenuous of the four sources studied.

\subsubsection{G25.411$+$00.105}

This maser ring has an inferred disk inclination of 47\degr\ from disk modeling, and therefore is expected to have an IR emission morphology as described in Scenario 3.

The 2.12\,$\mu$m map reveals a complex structure of NIR emission. Using a bright point source on the field, the 2.12~$\mu$m image was deconvolved to further enhance the resolution (to $\sim$80 mas). This is shown in Figure 6. The NIR peak labeled 3 corresponds most closely to the location of the maser ring center, with only a 1.9$\sigma$ (200 mas) offset, but this offset is in the direction expected from the toy model for this source (Figure 6). Consistent with Scenario 2 and the toy model, the bulk of the NIR emission seems to be distributed along the disk/outflow axis as defined by the fit to the maser ring (i.e., p.a. = 180\degr).

However, in the MIR images we see quite a different morphology (Figure 6). At both 8.6 and 18.3~$\mu$m, there is a well-defined peak, with an extended tongue of emission to the northwest. This main MIR peak is only 70 mas away from the NIR peak labeled source 3 and therefore assumed to be co-spatial.

The tongue of emission seen to the northwest of the peak in the MIR, is coincident with the NIR sources 6 and 7. In fact there are local weak peaks in the deconvolved MIR images that appear to be co-spatial with these NIR peaks. This MIR emission is not what would be expected for the morphology described in Scenario 3 or the toy model for this source, and therefore it does not seem to support the idea that the maser ring is in a circumstellar disk around source 3. However, further observations may be needed to determine if sources 6 and 7 are separate young stellar sources, or in any other way emission unrelated to the young stellar object with the maser ring. If they were found to be unrelated, this would give credence to the hypothesis that the masers here may be arising within a circumstellar disk.

Outflow observations do exist for this source but unfortunately not at sufficient resolution to clarify the situation. The outflow is seen best in the CO (2--1) maps of Beuther et al. (2002), which shows blue-shifted and red-shifted lobes with peaks situated at a position angle of $\sim$145$\degr$. The midpoint between these two peaks is $\sim$15$\arcsec$ to the northeast of the maser location, and therefore any YSO directly associated with the masers cannot be the main source driving the outflow. However, the CO contours have secondary peaks in both outflow lobes situated about $\sim$15$\arcsec$ to the west, with the blue-shifted CO outflow secondary peak coincident with the maser location and the MIR emission. There may therefore be two outflows located near each other in this region, but unresolved in the CO maps, one of which may be associated with the masers. The integrated HCO$+$ emission maps by Schenck et al. (2011), appear to only trace the blue-shifted lobe of the outflow, but interpretation is difficult with the limited resolution ($\sim$27$\arcsec$) of those maps. Further complicating the outflow picture is the fact that there is another bright NIR/MIR source 6$\arcsec$ to the north of the maser ring, which is likely a YSO unrelated to the masers. The position of this source is close to the midpoint between the secondary red-shifted CO peak and secondary blue-shifted peak, and could potentially be the source of the outflow, rather than the YSO associated with the maser ring.

In conclusion, while some evidence appears to support the disk hypothesis for this source, the MIR emission is not at all consistent with that picture.

\subsection{Results from SED model fitting}

The qualitative comparisons made so far between the expected IR morphologies and the observed IR morphologies are a good first-order test of the maser ring hypothesis. The next step is to see if the IR emission observed for each source quantitatively makes sense with respect to the maser ring hypothesis. A first cut at this can be done with existing YSO spectral energy distribution (SED) models, insofar as the change of the integrated flux density of an object as a function of wavelength will vary substantially as a function of MYSO geometries, including disk inclination and outflow cavity opening angle. Through the use of our observed flux densities and those available from data archives, we should be able to constrain the modeling enough to at least rule out certain geometries.

In order to derive and interpret the SEDs, we applied the model of young stellar objects presented by \citet{r07}\footnote{The tool available via http://caravan.astro.wisc.edu/protostars/}. First we searched for counterparts of the four targets in the following catalogues: 2MASS, {\it Spitzer} IRAC, MSX, and IRAS. For all four maser sources, the detected IRAS fluxes were used as upper limits in the SED models, due to the extremely coarse resolution of the IRAS data ($\sim$1$\arcmin$) and thus their potential for contamination. The 2MASS, IRAC, and MSX data were also taken with coarser resolutions than our Gemini NIR and MIR images (as mentioned in Sect.~1), however those same Gemini data were used to determine if these archival data would be used as upper limits or not.  The source IR fluxes and distances that we used in the SED modeling are listed in Table 3, and data used as upper limits are marked.

The model by \citet{r07} is based on the computed 20,071 YSO radiation transfer models at ten different viewing angles that span a large range of evolutionary stages and stellar masses. All of the main features of our toy model are formally integrated into these models; they use pre-main-sequence stars with different combinations of axis-symmetric circumstellar disks, infalling flattened envelopes, and outflow cavities under the assumption that stars form via accretion through the disk and envelope. The fitting procedure interpolates the model fluxes to the apertures used in measurements (the source sizes) scaling them to a given distance range of a source. It also allows the interstellar extinction, $A_V$, to be a free parameter in the fitting process, which we chose to be 0$<$$A_V$$<$100\,mag.

We present in Figure 7 the model fits of the SEDs. In Table 4 we list the derived physical properties of the modeled MYSOs, including stellar mass ($M_\star$), disk inclination ($i_{disk}$), disk outer radius ($R_{disk}$), outflow cavity opening angle ($\theta_{cavity}$), and total luminosity ($L_{tot}$). Several other properties are also given by the SED fits, however many of them are degenerate or not as important for the comparisons made here. For each source we examine the best 10 SED fits based on their $\chi^2$ values and list the best fitted values, the modes and medians for each parameter, and also give the minimum and maximum values of each parameter.  It can be seen in Table 4 that the central young stellar objects are indeed massive (8--15\,$M_{\odot}$ for the best fits), as expected for sources exciting methanol maser emission.

In Table 4, we also present the disk inclination and radius as derived from the maser ring geometry for each source. These can be directly compared to the MYSO geometries derived from the SED modeling to check for consistencies/inconsistencies of the maser disk hypothesis or our interpretations made so far based on the qualitative morphologies of the IR emission.

In the case of G23.389$+$00.185, we note a good agreement between the photometry from the high resolution MIR and NIR images and with the archive data. This translates into a fairly well constrained fit to the fluxes (Figure 7). This also can be seen in the similarities between the best fit, the mode fit, and the median fit in Table 4. The problem with this fit is that it seems to be quite different from what would be expected if the maser ring were tracing a disk. The maser ring geometry implies a disk inclination of 54$\degr$ and a radius of 425\,AU, however the models show that the most likely fits come from a more face-on geometry with a disk inclination of 18$\degr$ and a radius less than 100\,AU (and in most cases no disk at all). This adds further evidence to our claim from the discussion of the IR morphologies that the maser ring is not consistent with the disk hypothesis.

For G23.657$-$00.127, we again note a good agreement between the Gemini MIR and NIR photometry with archive data, which again translates into relatively well-constrained SED results. However, the results do exhibit a bimodal nature, with two groupings of tightly constrained parameter values that can be separated by disk inclination. For this source, therefore, we show the results of two top ten lists in Table 4. The first group of results are for the more face-on disk inclination, and the second for the more inclined disk results. It can be seen from the statistics of these two groups that there is very little spread in properties once the results are separated by the two disk inclinations. This is also evident in the SED plot for this source (Figure 7).

Interestingly, this dichotomy in the SED model results parallels the results from Section 3.3.1 based solely on the IR morphologies for this source. From the maser ring geometry alone, one would expect a nearly face on disk ($i=16\degr$). Recall that the MIR morphology by itself is consistent with a face-on geometry. Consistent with that picture, we find here that the best SED model is for a disk with a inclination of 18$\degr$. Conversely, however, recall that the NIR emission morphology is fan-like, and looks more like a more like a wide outflow cavity inclined to the line of sight. We see here that this geometry is more consistent with the majority of the top 20 SED modeling results which yield a disk with a high inclination (41--76$\degr$). Moreover, the SED results for an inclination of 18$\degr$ seem to create further problems for the maser ring hypothesis for this source, because for all of these fits (save the best fit) require diskless systems. In other words, the SED models that predict a low system inclination also predict an extremely small or diskless system, and therefore seem to contradict the maser disk hypothesis for this source.

For G24.634$-$00.324 the data did not constrain the models very well. Under the assumption that the IR morphology is due to an edge-on system, we used the combined flux of all emission on the Gemini fields in the SED modeling. The results all fell into three categories, the first being disk-only systems, the second being nearly face-on geometries, and the third being the rest of the results. We can, from the morphology in the high resolution IR images alone, ignore all results from fits that imply a face-on geometry. Second, as we have discussed in Section 3.2, we have good reason to believe that these sources are very young MYSOs still embedded in their accretion envelopes, so we also ignore the SED results that are consistent with disk-only systems. The top 10 remaining fits are shown in Table 4 and Figure 7. The parameters are consistent with a relatively inclined system ($i=41-76\degr$) and therefore lend some credence to the belief that the MIR emission seen in the Gemini images comes from a ``silhouette-like'' disk. This however, would also confirm that the maser ring is elongated at the wrong angle to be tracing the disk in the ``dark lane'' as seen in the IR. If we just use the Gemini fluxes for source 2 and use the Spitzer and MSX data as upper limits (since they do not resolve source 1 and 2) the SED fits imply a central stellar mass less than 8 $M_{\odot}$. This seems to again strengthen the case for the IR emission morphology being from a MIR ``silhouette-like'' disk (rather than a cluster of three individual YSOs) and that the IR images are inconsistent with the maser ring hypothesis.

In the case of G25.411$+$00.105, we assume that all of the flux in the Gemini images is from the outflow cavity of a single source. However, because of the close proximity of another bright IR source $\sim$6$\arcsec$ to the north, the $\sim$18$\arcsec$ resolution MSX data have been used as upper limits (Table 3) for these fits. The results from the fits are mixed; some parameters seem well-constrained (i.e. $i_{disk}$ and $M_{\star}$), but others are not. Interestingly, all fits yield an inclined disk geometry that is consistent with the maser ring inclination. However, the derived cavity opening angle ($\theta$$_{cavity}$ = 6$\degr$; see Table 4) may be too narrow to explain the IR emission as all coming from the blue-shifted outflow cavity. Furthermore, the maser ring is 550\,AU in radius, and the median disk size of the fits is only a fourth of that, with the best fit being only 12\,AU. This casts further doubt on the interpretation that the maser ring is in a circumstellar disk around this source.

\subsection{Do methanol maser rings trace circumstellar disks?}

Of the four methanol maser rings we observed in the infrared, none convincingly displayed the characteristics expected if the maser rings are indeed tracing circumstellar disks. In the case of G23.657-00.127, the maser ring is nearly circular, indicating that we should be looking almost straight down the outflow cavity (i.e., Scenario 1). However, we see that the  2 $\mu$m morphology is fan-shaped as one would expect for the reflected light from a outflow cavity with a much more inclined geometry. The majority of SED model fits to the infrared photometry of this source seem to corroborate this more inclined disk geometry. In the case of G24.634-00.324, the maser ring is highly elliptical, indicating that we should be seeing a more edge-on disk geometry, and perhaps a silhouette-like infrared morphology where the masers reside in the ``dark lane'' of the infrared emission (i.e., Scenario 3). Consistent with the results from the SED modeling, we do indeed see this type of morphology in the MIR however the angle of the ``silhouette disk'' is almost 90$^{\circ}$ from the expected disk/outflow axis as given by the maser ring geometry. For G23.389+00.185, the maser ring is slightly elliptical, indicating that we will likely only see the disk face/outflow cavity facing towards us (i.e. Scenario 2). However, the maser ring center is significantly offset from the IR peak and at an orientation that is not consistent with emission from a disk/outflow as derived from the maser ring geometry. Finally, for G25.411+00.105 which also has a maser ring with a slight ellipticity (i.e. Scenario 2), has NIR emission consistent with scattering/reflection off of a extended outflow cavity and at the right orientation as inferred from the maser ring geometry. However, the MIR emission is elongated at an angle 130\degr different from this, and has a morphology that casts doubt on the validity of the conclusion that would be drawn from just the NIR emission and maser ring geometry alone. Furthermore, SED models have a hard time reproducing the large disk size implied by the maser ring geometry.

In summary, none of the four sources observed give compelling evidence in support of the hypothesis that maser rings may arise from circumstellar disks. Moreover, the two clearest examples of the maser ring morphology in our sample, G23.657-00.127 and G23.389+00.185, seem to have fairly good observational evidence that contradicts such claims. Therefore the conclusion from these observations is that the infrared emission from these sources does not seem to support the scenario where methanol maser rings trace circumstellar disks around young massive stars.

If maser rings are not tracing disks, what else could explain these ring structures? As discussed in the introduction, the scenarios that have been proposed to generally explain 6.7\,GHz methanol maser emission are disks, outflow and shocks. If the maser rings are not in disks, then they may be coming from shocks within outflows. Ellipses are created geometrically from the cross-section of a cone and a plane. Given the conical geometry of the outflow cavity, a quasi-planar accretion shock emanating from the (proto)star/disk could travel down the outflow cavity walls and cause a geometrically ring-like shock. This shock would liberate methanol from the dust, which could then be radiatively pumped to produce the maser emission. Though in principal this is a possibility, our data here cannot adequately address this hypothesis. The outflow cavities of MYSOs can be heated by MIR emission out to 10s of thousands of AU (De Buizer 2006), and therefore the ring produced in this way could be located anywhere along the outflow cavity. Furthermore, due to possible differences in density on opposite sides of an outflow cavity's walls, the propagation of shock down the cavity can skew the ring so that the minor axis of the ellipse does not point back to the central (proto)star/disk. This means that the location of the IR emission with respect to the maser ring center and with respect to the maser ring geometry are not well-constrained, unlike with the disk hypothesis. Therefore, with only the data we have here we cannot truly test the outflow shock hypothesis.

It seems the obvious way forward to solve the mystery of what causes the maser rings will be to employ ALMA to directly image the disks at high spatial resolution around these sources. Given directly observed disk geometries and their location with respect to the maser emission, one may finally be able to disentangle the true nature of methanol maser rings.

\section{Conclusions}

The goals of this paper were to resolve the NIR and MIR emission at the locations of four methanol maser rings to verify whether or not their IR morphologies are consistent with the hypothesis that methanol maser rings are tracing circumstellar disks, and to derive physical properties and geometries for their associated massive young stellar objects through SED model fitting.

In this article we discussed the assumed distribution of circumstellar material around such young and massive accreting (proto)stars, and what infrared emission geometries would be expected for different disk/outflow orientations. For the four targets we observed, we compared the expected infrared geometries (as inferred from the properties of the maser rings) to actual high spatial resolution near-infrared and mid-infrared images. We find that the observed infrared emission geometries are not consistent with the hypothesis that the masers are residing in circumstellar disks.

Using SED model fitting, we found that the emission from the infrared counterparts for all methanol masers ring distributions are indeed consistent with massive young stellar objects with masses above 8\,M$_{\sun}$. Furthermore, we find that in most cases the geometries allowed by the SED model fits corroborate the negative results from the observed infrared morphologies, casting further doubt on the hypothesis that the methanol maser rings in these four cases arise from within circumstellar disks.
\\

\acknowledgments
A.B.\ and M.S.\ acknowledge support by the Polish Ministry of Science and Higher Education through grant N203 386937. A.B.\ acknowledges also support by the Nicolaus Copernicus University grant 378-A (2010). We thank to Dr.~Eric Greisen from NRAO and Dr.~Jagadheep Pandian from University of Hawaii for fruitful discussions.

Based on observations obtained at the Gemini Observatory, which is operated by the Association of Universities for Research in Astronomy, Inc., under a cooperative agreement with the NSF on behalf of the Gemini partnership: the National Science Foundation (United States), the Science and Technology Facilities Council (United Kingdom), the National Research Council (Canada), CONICYT (Chile), the Australian Research Council (Australia), Minist\'{e}rio da Ci\^{e}ncia e Tecnologia e Inova\c{c}\~{a}o (Brazil) and Ministerio de Ciencia, Tecnolog\'{\i}a e Innovaci\'{o}n Productiva (Argentina).

This research has made use of the NASA/IPAC Infrared Science Archive, which is operated by the Jet Propulsion Laboratory, California Institute of Technology, under contract with the National Aeronautics and Space Administration.

{\it Facilities:} \facility{Gemini North (NIRI/Altair), Gemini South (T-ReCS)}.

\appendix
\section{Additional Information on Sources}
During the study of the sources in this article, interesting results were gathered that do not have any direct relationship to the main goals of the paper. However, this information is likely to be of use to those who study these sources in-depth. Here below we summarize auxiliary information we have for two sources in this study.

{\bf G23.389$+$00.185.}  The MIR emission towards this source is the brightest among five targets. We note this methanol ring was also the brightest maser in the selected sample of methanol sources with S$_{\rm p}$=21.55\,Jy\,beam$^{-1}$ \citep{b09}.  We also found the 22\,GHz water masers towards this source (the only detection of the four in the sample) with a distribution along a position angle of 45\degr\, from the major axis of the fitted ellipse to the 6.7\,GHz methanol spots \citep{b10}. However, their position uncertainties are of order 0\farcs15.

Interestingly, there is a bright jet of emission seen in this image, that can be better seen in a unsharp mask of the data (see Figure 2). It appears to point straight back to source 2. Jets from outflowing massive young stellar objects have be observed in infrared continuum \citep{d06}. However, this jet-like structure is not seen in the MIR, thus indicating that if it is a jet it must be line emission, not continuum. The $K^{\prime}$ filter of NIRI is centered on the 2.122~$\mu$m $\nu = 0-1$ S(1) line of H$_2$, a well-known outflow indicator. This therefore likely a jet coming from source 2.

{\bf G25.411$+$00.105.}  The brightest NIR peak, source 1, has no MIR counterpart, and there are many other peaks identified in the NIR that also have no MIR counterparts. This means that, like the fan-like extended structure of G23.657$-$00.127, the NIR emission could be scattered emission, or alternatively it may line emission, as discussed for G23.389$+$00.185 above.


\clearpage

\begin{deluxetable}{rccccccc}
\tabletypesize{\tiny}
\tablecaption{Maser Fields and the Positions of Detected Infrared Sources}
\tablewidth{0pt}
\tablehead{
\colhead{} & \multicolumn{3}{c}{NIR (2.12\,$\mu$m)} & & \multicolumn{3}{c}{MIR (8.6\,$\mu$m)} \\
\cline{2-4} \cline{6-8} \\
\colhead{} & \multicolumn{2}{c}{Peak Position (J2000)} & \colhead{1$\sigma$ Error} & & \multicolumn{2}{c}{Peak Position (J2000)} & \colhead{1$\sigma$ Error} \\
\colhead{Maser Field : IR Source} &\colhead{$\alpha$} &\colhead{$\delta$} &\colhead{($\Delta$$\alpha$,$\Delta$$\delta$)} & &\colhead{$\alpha$} &\colhead{$\delta$} &\colhead{($\Delta$$\alpha$,$\Delta$$\delta$)}\\
\colhead{} &\colhead{(h m s)} &\colhead{(\degr\, \arcmin\, \arcsec)} & \colhead{(mas,mas)} & &\colhead{(h m s)} &\colhead{(\degr\, \arcmin\, \arcsec)} & \colhead{(mas,mas)}
}
\startdata
G23.389$+$00.185:BDS12 1  & 18 33 14.323 & $-$08 23 57.30 & 155,15 & & 18 33 14.327 & $-$08 23 57.29 & 60,60\\
BDS12 2 & 18 33 14.417 & $-$08 23 55.76 &155,15 & & 18 33 14.417 & $-$08 23 55.78 & 60,60 \\
G23.657$-$00.127:BDS12 1 & 18 34 51.561 & $-$08 18 21.52 & 106,157 & & 18 34 51.556 & $-$08 18 21.51 & 60,60\\
G24.634$-$00.324:BDS12 1 &\nodata &\nodata &\nodata & & 18 37 22.683 & $-$07 31 41.58 &60,60 \\
BDS12 2 &\nodata &\nodata &\nodata & & 18 37 22.713 & $-$07 31 42.33  &60,60 \\
BDS12 3 &\nodata &\nodata &\nodata & & 18 37 22.659 & $-$07 31 42.68 &60,60 \\
G25.411$+$00.105:BDS12 1 & 18 37 16.920 & $-$06 38 31.00 &114,96  & &\nodata &\nodata &\nodata \\
BDS12 2 & 18 37 16.908 & $-$06 38 30.74 &114,96  & &\nodata &\nodata &\nodata\\
BDS12 3 & 18 37 16.917 & $-$06 38 30.62 &114,96  & & 18 37 16.916 & $-$06 38 30.58 & 60,60 \\
BDS12 4 & 18 37 16.922 & $-$06 38 31.57 &114,96  & &\nodata &\nodata & \nodata\\
BDS12 5 & 18 37 16.957 & $-$06 38 30.21 &114,96  & &\nodata &\nodata & \nodata\\
BDS12 6 & 18 37 16.906 & $-$06 38 30.48 &114,96  & &\nodata &\nodata & \nodata\\
BDS12 7 & 18 37 16.898 & $-$06 38 30.27 &114,96  & & 18 37 16.897 & $-$06 38 30.29 &60,60 \\
BDS12 8 & 18 37 16.866 & $-$06 38 30.58 &114,96  & &\nodata &\nodata & \nodata\\
BDS12 9 & 18 37 16.875 & $-$06 38 30.05 &114,96  & &\nodata &\nodata & \nodata
\enddata
\label{table:1}
\end{deluxetable}

\begin{deluxetable}{rcccc}
\tabletypesize{\scriptsize}
\tablecaption{Integrated Flux Densities of Detected Infrared Sources}
\tablewidth{0pt}
\tablehead{
\colhead{} & \multicolumn{3}{c}{F$_{\lambda,\rm int}$} & \\
\cline{2-4} & & & &\colhead{Aperture}\\
\colhead{Infrared Source Name} &\colhead{2.12\,$\mu$m} &\colhead{8.6\,$\mu$m} &\colhead{18.3\,$\mu$m} &\colhead{Radius} \\
\colhead{} &\colhead{(mJy)} &\colhead{(Jy)} & \colhead{(Jy)} &\colhead{($\arcsec$)} 
}
\startdata
G23.389$+$00.185:BDS12 1   &294  &21.9 &45.7 & 1.6 \\
                 BDS12 2   &1.41  &0.12  &1.21  & 0.4 \\
G23.657$-$00.127:BDS12 1   &81.0   &7.04  &25.7 & 2.6 \\
G24.634$-$00.324:BDS12 1   &\nodata   &2.41  &4.67  & 0.7 \\
                 BDS12 2   &\nodata   &0.29  &1.97  & 0.5 \\
                 BDS12 3   &\nodata   &0.04  &$<$0.79 & 0.4\\
                Combined   &\nodata   &2.91  &7.99  & 1.5 \\
G25.411$+$00.105:BDS12 1   &1.83 &       &       & 0.2 \\
                 BDS12 2   &0.92 &       &       & 1.5 \\
                 BDS12 3   &0.28 &0.28  &1.91  & 0.4\\
                 BDS12 4   &0.21 &       &       & 0.2 \\
                 BDS12 5   &0.04 &       &       & 0.2 \\
                 BDS12 6   &0.15 &       &       & 0.1 \\
                 BDS12 7   &0.14 &0.05  &0.20  & 0.1 \\
                 BDS12 8   &0.17 &       &       & 0.2 \\
                 BDS12 9   &0.05 &       &       & 0.1 \\
                Combined   &6.32 &0.36  &2.02  & 1.2 
\enddata
\tablecomments{All flux densities are background subtracted and derived using circular aperture photometry. The aperture radius given for each source was used for all wavelengths. ``Combined'' fluxes are for an aperture enclosing all emission in the region around the masers. Flux errors (1$\sigma$) are 5\% for 2.12~$\mu$m, 8\% for 8.6~$\mu$m, and 15\% for 18.3~$\mu$m. No 2.12\,$\mu$m data were taken for G24.634-00.324. Blank entries denote no detection. Upper limit for detection of a point source is 2.1~mJy at 8.6~$\mu$m, and 36.7~mJy for 18.3~$\mu$m.}
\label{table:2}
\end{deluxetable}

\begin{deluxetable}{lllcccc}
\tabletypesize{\scriptsize}
\tablecaption{Auxiliary Inputs to the SED models$^a$}
\tablewidth{0pt}
\tablehead{
 & \multicolumn{2}{c}{Band} & \colhead{G23.389$+$00.185}  & \colhead{G23.657$-$00.127}  &  \colhead{G24.634$-$00.324} & \colhead{G25.411$+$00.105}  \\
 & & & \colhead{BDS12 1}&\colhead{BDS12 1} & \colhead{Combined} & \colhead{Combined} \\
 & & \colhead{($\mu$m)}  & \colhead{(Jy)} & \colhead{(Jy)} & \colhead{(Jy)} & \colhead{(Jy)} 
}
\startdata
2MASS & J& 1.25& 0.0007$\pm$0.00007& 0.0005$\pm$0.00005 & 0.000767$\pm$0.00007& 0.0002$\pm$0.0002 \\
      & H& 1.65& 0.021$\pm$0.002& 0.01$\pm$0.001 & 0.005$\pm$0.0005&0.0006$\pm$0.00006 \\
      & K& 2.17& 0.28$\pm$0.028& 0.07$\pm$0.007 & 0.064$\pm$0.006&0.0048$\pm$0.00048 \\
Spitzer & IRAC [1] &3.6& 2.08$\pm$0.02 & 1.2$\pm$0.02 &0.58$\pm$0.06 & 0.09$\pm$0.009 \\
        & IRAC [2] &4.5& 3.45$\pm$0.1 & 2.1$\pm$0.1 & 1.3$\pm$0.13 & 0.4$\pm$0.04 \\
        & IRAC [3] &5.8& 14.0$\pm$ 0.4  & 7.4$\pm$0.4 & 3.1$\pm$0.31 & 0.85$\pm$0.085 \\
        & IRAC [4] &8.0& --$^b$           & 7.3$\pm$1 & 3.0$\pm$0.30 & 1.0$\pm$0.1 \\
MSX   & A &8.28& 23.47$\pm$2.4& 8.518$\pm$0.85 & 4.50 &2.3$\downarrow$ \\
      & C &12.13& 39.00$\pm$3.9& 17.27$\pm$1.7 & 6.70 &3.4$\downarrow$ \\
      & D &14.64& 50.99$\pm$5.1& 30.45$\pm$3.05 & 8.78 & 4.7$\downarrow$ \\
      & E &21.34& 54.71$\pm$5.5& 43.87$\pm$3.4 & 13.47 & 8.4$\downarrow$ \\
IRAS  & &12 & 35.8$\downarrow$ &16.0$\downarrow$ &10.3$\downarrow$ & 3.5$\downarrow$ \\
      & &25 & 76.4$\downarrow$ &70.3$\downarrow$&25.0$\downarrow$ & 17.5$\downarrow$ \\
      & &60 & 250$\downarrow$  &383$\downarrow$ &91.8$\downarrow$ & 110$\downarrow$ \\
      & &100 & 258$\downarrow$  &402$\downarrow$ &258$\downarrow$& 567$\downarrow$ \\
\hline\\ [-1.5ex]
D$_{adopted}$$^c$  & (kpc) & & 4.47  & 3.19$^d$   &  3.0 & 5.35

\enddata
\tablecomments{All sources have IRAS and MSX detections, however if these measured flux densities are used as upper limits in the SED modeling, this is denoted by a $\downarrow$ following those values in the table.}
\tablenotetext{a}{Also included in the modeling were the relevant source flux densities from Gemini observations in Table 2.}
\tablenotetext{b}{Source is saturated in this image.}
\tablenotetext{c}{Assumed distance to source adopted in this paper. With the exception of G23.657-00.127, these are the near kinematic distances for the sources based on the calculations from Reid et al. (2009). For SED models, the distance range used was this value $\pm$10\%.}
\tablenotetext{d}{Distance based on the trigonometric parallax (Bartkiewicz et al. 2008).}
\label{table:3}
\end{deluxetable}

\begin{deluxetable}{lccccc}
\tabletypesize{\scriptsize}
\tablecaption{Derived Physical Parameters from SED Models}
\tablewidth{0pt}
\tablehead{
\colhead{} & \colhead{$i_{disk}$} & \colhead{$R_{disk}$}  & \colhead{$M_{\star}$}   & \colhead{$L_{tot}$}  & \colhead{${\theta}_{cavity}$} \\
\colhead{} & \colhead{($\degr$)}    & \colhead{(AU)}        & \colhead{($M_{\odot}$)} & \colhead{(10$^3$\,$L_{\odot}$)} & \colhead{($\degr$)}
}
\startdata
\multicolumn{6}{c}{G23.389$+$00.185:BDS12 1}    \\	
\multicolumn{6}{c}{Maser ring: $i$=54$\degr$; $R$=425\,AU}	\\
\hline \\ [-1.5ex]
Best		&18			&58				&15			&15			 &5		 \\	
Mode		&18			&48				&14			&17			 &9		 \\	
Median		&18			&36				&14			&12			 &6		 \\	
Range		&18--57		&0$^a$--93		&13--21		&9--17		 &4--25	 \\
\hline \\ [-1.5ex]
\multicolumn{6}{c}{G23.657$-$00.127:BDS12 1}\\	
\multicolumn{6}{c}{Maser ring: $i$ = 16$\degr$; $R$ = 425\,AU}  \\
\hline \\ [-1.5ex]
Best$^b$		&18			&21				&12			&6			 &7	 \\		
Mode$^b$		&18			&0$^a$			&--			&--			 &--	 \\		
Median$^b$		&18			&7				&13			&5			 &6	 \\		
Range$^b$ 		&18--18		&0$^a$--21		&11--16		&4--7	   	 &5--8\\			
\\
Best$^c$		&70			&179		&11			&8			 &49	 \\		
Mode$^c$		&76			&179		&11			&8			 &49	 \\		
Median$^c$	&66			&179		&11			&8			&49	 \\		
Range$^c$ 	&41--76		&62--179	&11--12		&8--11		 &45--55	 \\	
\hline \\ [-1.5ex]
\multicolumn{6}{c}{G24.634$-$00.324:Combined} \\		
\multicolumn{6}{c}{Maser ring: $i$ = 71$\degr$; $R$ = 135\,AU}  \\
\hline \\ [-1.5ex]
Best				&57			&191		&9			&4			 &50		 \\	
Mode$^d$		&63			&191		&9			&4			 &50		 \\	
Median$^d$	&63			&217		&8			&3			&34		 \\	
Range$^d$		&41--76		&65--1880	&7--9		&3--4		 &24--50	\\	
\hline \\ [-1.5ex]
\multicolumn{6}{c}{G25.411$+$00.105:Combined}	\\	
\multicolumn{6}{c}{Maser ring: $i$ = 47$\degr$; $R$ = 550\,AU}  \\
\hline \\ [-1.5ex]
Best			&32			&12			&9			&1			 &6			 \\
Mode$^e$		&32			&--			&10			&--			 &--			 \\
Median$^e$		&32			&157		&8			&2			 &6			 \\
Range$^e$		&32--57		&7--782		&8--11		&1--8		 &2--34		 
\enddata
\tablenotetext{a}{Means diskless.}
\tablenotetext{b}{Results are bimodal. Two top tens tabulated. Best overall fit comes from this group of results and implies a very small disk. Four more of these top 10 fits are diskless.}
\tablenotetext{c}{Results are bimodal. Not including the best overall fit and the low-inclination cases.}
\tablenotetext{d}{Results are trimodal but only showing model fits consistent with observations. Therefore, not showing results
  for those with disk-only fits and/or with face-on disk inclinations.}
\tablenotetext{e}{One of the top ten fits was for a 1.7\,$M_{\odot}$ source and was removed.}
\label{table:4}
\end{deluxetable}


\clearpage

\begin{figure*}
\centering
\includegraphics[scale=0.8]{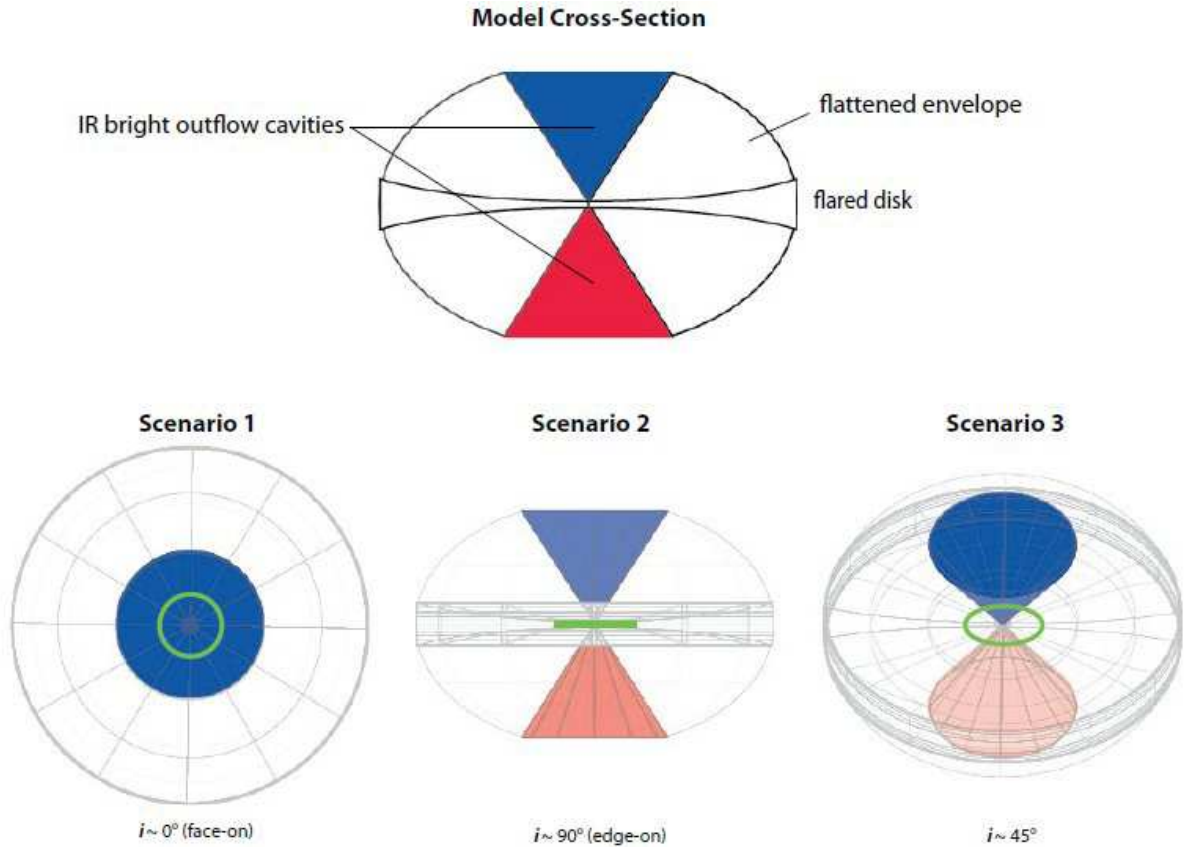}
 \caption{A cross-section view of the toy model for a MYSO and 3D renderings of the model at disk/system inclinations given by the scenarios outlined in Section 3.2. In the cross-section view, the IR bright cavities are shown in blue and red, for what will be the blue-shifted and red-shifted outflow cavities, respectively, in the 3D renderings below it. The envelope and flared accretion disk are labeled. In the three 3D renderings of this model for the three inclination scenarios, the disk and envelope are not 100\% opaque, and thus the lightening of the red and blue cavities reflect the effects of obscuration by the disk and envelope. In reality, the more optically thick the disk and envelope, the more attenuated the emission from the cavities, especially the red-shifted one. The maser rings are represented by the green circles, demonstrating their location with respect to the IR emission as a function of system inclination.
}
\label{fig1}
\end{figure*}

\clearpage

\begin{figure*}
\centering
\includegraphics[scale=0.7]{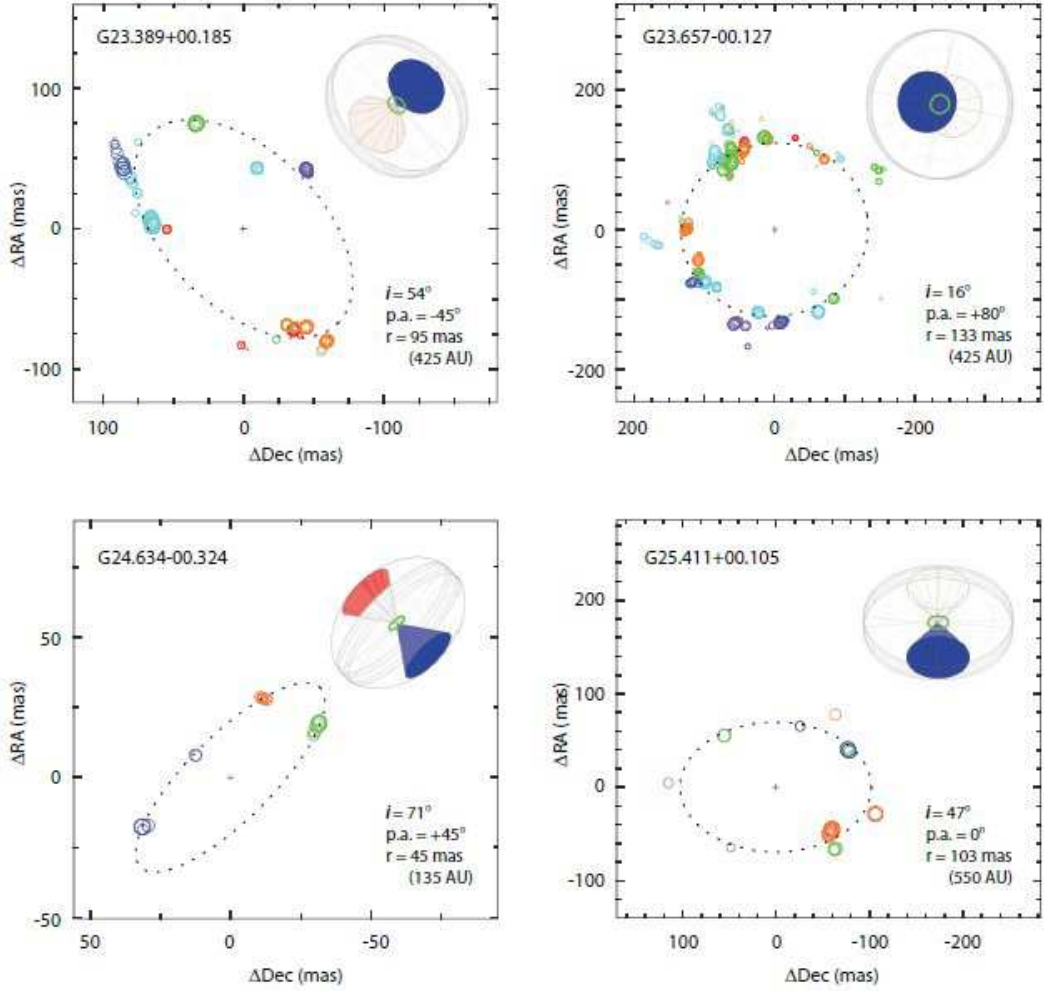}
 \caption{The four methanol maser rings in the sample with their implied disk properties, assuming the masers arise from within circumstellar disks. In each panel the multi-colored circles represent the individual maser spots, with colors indicating line of sight velocity of each spot (see Bartkiewicz et al. 2009 for these values). The dotted line is the best fitted ellipse to the maser spot distribution, with the cross demarcating the center of the ellipse. A toy model (as described in Figure 1) is displayed in the upper right corner of each panel, using the properties (from Bartkiewicz et al. 2009) given in the lower right corner. The properties given are disk inclination, position angle of the disk rotational symmetry axis (or the outflow axis) from north to east, and the radius of the disk given by the semi-major axis of the maser ring fit. The blue cone of the disk model shows the assumed blue-shifted outflow side of the MYSO, and the red cone the red-shifted side. The green ellipse shows the expected location of the maser ring if it is in the disk.
}
\label{fig2}
\end{figure*}

\clearpage

\begin{figure*}
\centering
\includegraphics[scale=0.88]{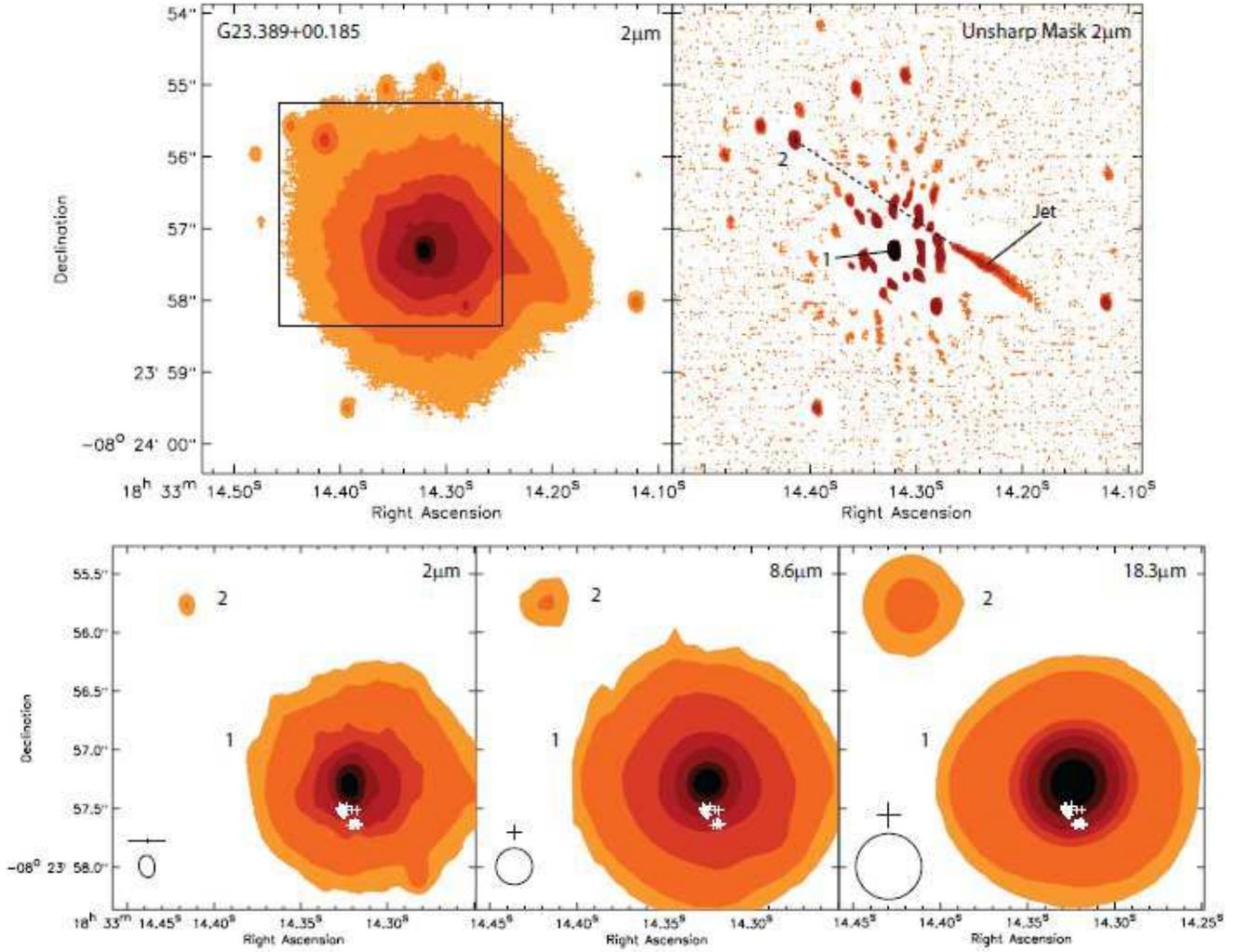}
 \caption{The NIR and MIR emission registered towards G23.389$+$00.185. The top left box shows a wider field view of the 2\,$\mu$m image, and a unsharp mask of the same image to its right. We detect a jet-like feature in the NIR that appears to be coming from source 2 (see appendix). The box in the top left panel is the field of view of the three panels below. The 6.7\,GHz methanol masers from Figure 2 are represented by white crosses. The $\pm$1$\sigma$ astrometric errors in Table 1 are shown for each wavelength with the black cross in the lower left of each panel, along with the effective resolution (azimuthally averaged FWHM of the PSF stars) as the circle. For this source only, the 2\,$\mu$m image was sufficiently elongated that we show the 50\% level of an actual star on the field in the bottom left panel. The colors on the bottom three panels trace the emission levels as follows: 4, 7, 20, 50, 100, 300, 900~$\mu$Jy~pixel$^{-1}$ (at 2~$\mu$m); 2, 4, 10, 60, 150, 400, 600~mJy~pixel$^{-1}$ (at 8.6~$\mu$m); and 8, 20, 60, 120, 200, 400, 600~mJy~pixel$^{-1}$ (at 18.3~$\mu$m).}
\label{fig3}
\end{figure*}

\clearpage

\begin{figure*}
\centering
\includegraphics[scale=0.9]{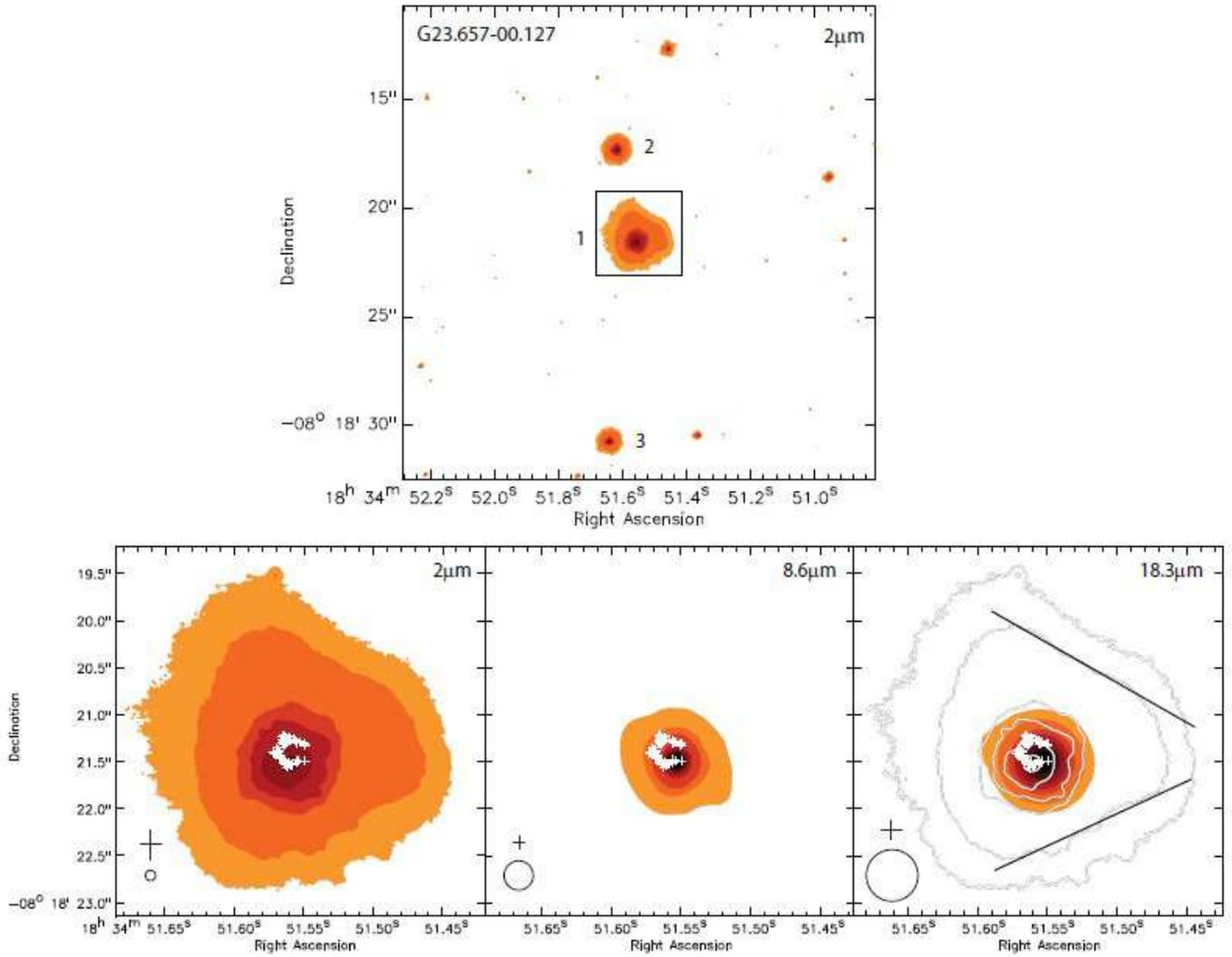}
 \caption{The NIR and MIR emission registered towards  G23.657$-$00.127.  In the right panel we overlay the 2\,$\mu$m contours on the 18.3\,$\mu$m image. Also plotted are two lines to demonstrate how straight and flat the sides of the NIR emission are. Symbols are the same as in Figure 3. The colors on the bottom three panels trace the emission levels as follows: 0.1, 1, 5, 10, 20, 200, 800~$\mu$Jy~pixel$^{-1}$ (at 2~$\mu$m);  5, 25, 50, 150, 200, 250, 300~mJy (at 8.6~$\mu$m); and 35, 75, 150, 200, 250, 300, 350~mJy~pixel$^{-1}$ (at 18.3~$\mu$m).
}
\label{fig4}
\end{figure*}

\clearpage

\begin{figure*}
\centering
\includegraphics[scale=0.65]{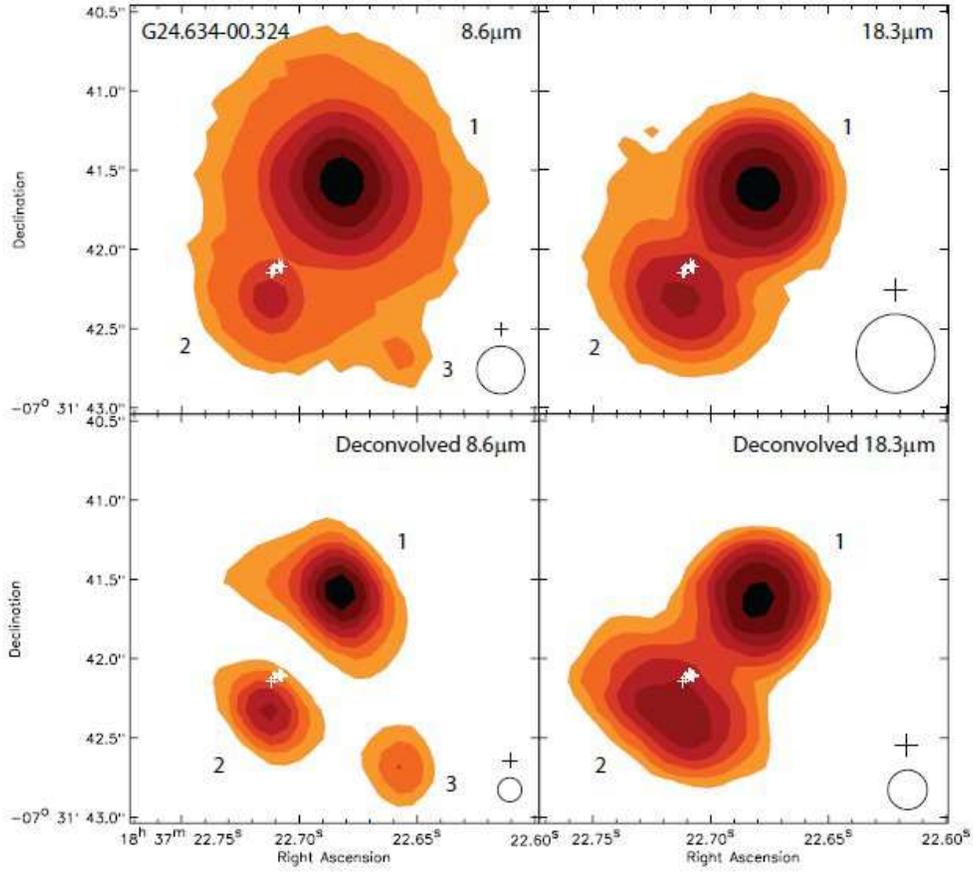}
 \caption{The MIR emission towards G24.634$-$00.324. The two top panels show the MIR images with their native resolution, whereas the bottom two panels show the same images deconvolved. Symbols are the same as in Figure 3. The colors trace the emission levels as follows: 0.8, 1.3, 4.7, 7.0, 14, 25, 60~mJy~pixel$^{-1}$ (for 8.6~$\mu$m); 8.0, 13, 18, 23, 32, 50, 70~mJy~pixel$^{-1}$ (for 18.3~$\mu$m); 0.5, 1.3, 3.7, 10, 27, 74, 200~mJy~pixel$^{-1}$ (for deconvolved 8.6~$\mu$m); and 2.0, 4.5, 10, 22, 50, 112, 230~mJy~pixel$^{-1}$ (for deconvolved 18.3~$\mu$m).
}
\label{fig5}
\end{figure*}

\clearpage

\begin{figure*}
\centering
\includegraphics[scale=0.9]{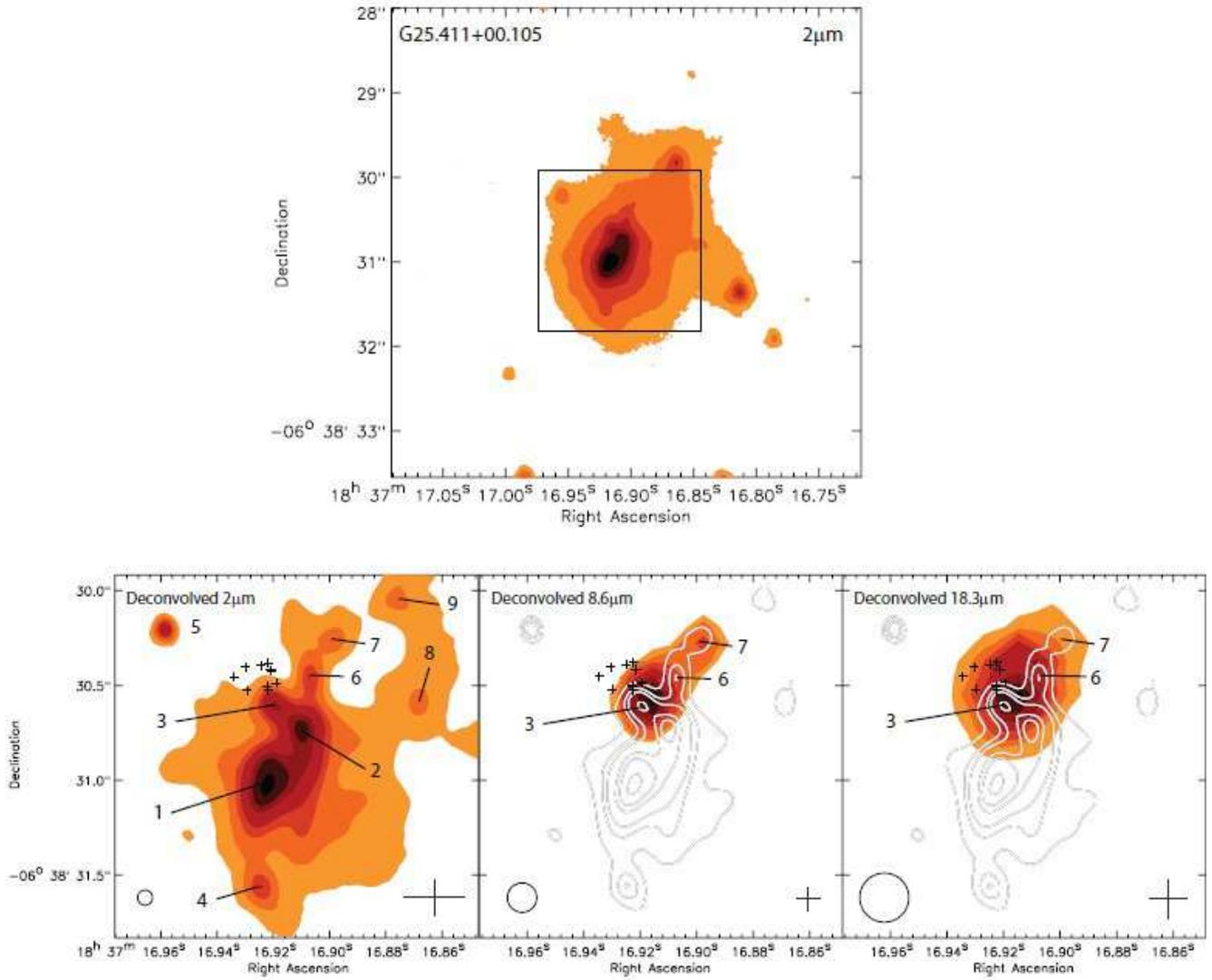}
 \caption{The NIR and MIR emission registered towards G25.411$+$00.105. The upper panel shows a wider field view of the 2\,$\mu$m image at the natural resolution using the AO system, with the box denoting the field in the lower panels. The lower panels show the deconvolved images at 2, 8.6, and 18.3\,$\mu$m. The 8.6 and 18.3\,$\mu$m images have the 2\,$\mu$m contours overlaid. Black crosses denote the masers for this source as given in Figure 2. All other symbols are as described in Figure 3. The colors on the bottom three panels trace the emission levels as follows:  0.3, 0.6, 1, 2, 4, 5, 7~$\mu$Jy (at 2~$\mu$m); 1, 2, 5, 10, 20, 35, 50~mJy (at 8.6~$\mu$m); and 2, 10, 25, 50, 90, 130, 170~mJy (at 18.3~$\mu$m).
}
\label{fig6}
\end{figure*}

\clearpage

\begin{figure*}
\centering
\includegraphics[scale=0.8]{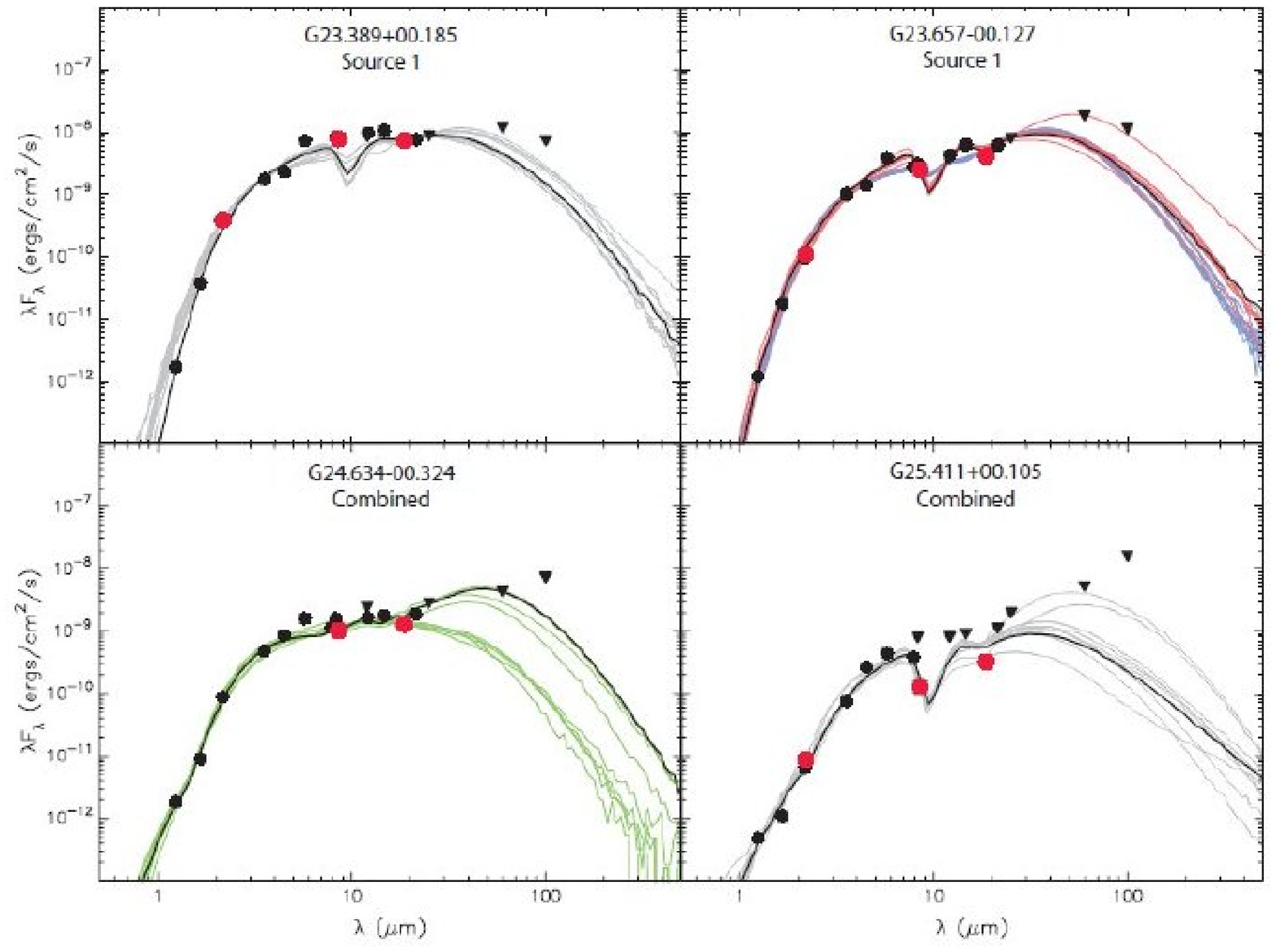}
 \caption{Spectral energy distributions of the continuum emission towards the four maser rings based on the data in Table 4. The filled circles and triangles show the input fluxes and upper limits, respectively. Red circles represent the Gemini data presented here. The black line shows the best fit, and the gray lines show the subsequent 9 good fits. For G23.657-00.127 the results were bimodal (see Section 3.4) and the red fits correspond to the group of fits with $i_{disk}$$\sim$18$\degr$, and the blue correspond to the group of fits with $i_{disk}$$\sim$70$\degr$. The green fits for G24.634-00.324 denote that these resultant fits are not the overall top ten best fits, but the ten best model fits after some selection criteria were employed (see Section 3.4).
}
\label{fig7}
\end{figure*}


\begin{thebibliography}{99}
\bibitem[Alvarez et al. (2004)]{} Alvarez, C., Hoare, M., \& Lucas, P.\ 2004, \aap, 419, 203
\bibitem[\protect\citeauthoryear{Bartkiewicz et al.}{2005}]{b05}
    Bartkiewicz, A., Szymczak, M., \& van Langevelde, H.~J.\ 2005, \aap, 442, L61
\bibitem[\protect\citeauthoryear{Bartkiewicz et al.}{2008}]{b08} Bartkiewicz, A., Brunthaler, A., Szymczak, M., van Langevelde, H.~J. \& Reid, M.~J.\ 2008, \aap, 490, 787
\bibitem[\protect\citeauthoryear{Bartkiewicz et al.}{2009}]{b09}
    Bartkiewicz, A., Szymczak, M., van Langevelde, H.~J., Richards, A.~M.~S., \& Pihlstr{\"o}m, Y.~M.\ 2009, \aap, 502, 155
\bibitem[\protect\citeauthoryear{Bartkiewicz et al.}{2011}]{b10}
    Bartkiewicz, A., Szymczak, M., Pihlstr{\"o}m, Y.~M., et al.\ 2011, \aap, 525, A120
\bibitem[Beuther et al.(2002)]{} Beuther, H., Schilke, P., Sridharan, T.~K., et al.\ 2002, \aap, 383, 892
\bibitem[\protect\citeauthoryear{Beuther \& Shepherd}{2005}]{bs09} Beuther, H. \& Shepherd, D.\ 2005, Cores to Clusters: Star Formation with Next Generation Telescopes, 105
\bibitem[\protect\citeauthoryear{Beuther et al.}{2009}]{be09} Beuther, H., Walsh,A.~J., \& Longmore, S.~N.\ 2009, \apjs, 184, 366
\bibitem[\protect\citeauthoryear{Beuther et al.}{2010}]{be10} Beuther, H., Henning, T., Linz, H., et al.\ 2010, \aap, 518, L78
\bibitem[Breen et al.(2010)]{} Breen, S.~L., Ellingsen, S.~P., Caswell, J.~L., \& Lewis, B.~E.\ 2010, \mnras, 401,
\bibitem[\protect\citeauthoryear{Caswell et al.}{1995}]{c95} Caswell, J.~L., Vaile, R.~A., Ellingsen, S.~P., Whiteoak, J.~B., \& Norris, R.~P.\ 1995, \mnras, 272, 96
\bibitem[Chini et al. (2004)]{} Chini, R., Hoffmeister, V., Kimeswenger, S., et al.\ 2004, \textit{Nature}, 429, 155
\bibitem[Cohen et al.(1999)]{} Cohen, M., Walker, R.~G., Carter, B., et al.\ 1999, \aj, 117, 1864
\bibitem[Cotera et al. (2001)]{} Cotera, A.~S., Whitney, B.~A., Young, E., et al.\ 2001, \apj, 556, 958
\bibitem[Cragg et al. (2005)]{} Cragg, D.~M., Sobolev, A.~M., \& Godfrey, P.~D.\ 2002, \mnras, 331, 521
\bibitem[\protect\citeauthoryear{Cutri et al.}{2003}]{c03} Cutri, R.~M., Skrutskie, M.~F., van Dyk, S., et al.\ 2003, The IRSA 2MASS All-Sky Point Source Catalog, NASA/IPAC Infrared Science Archive, http://irsa.ipac.caltech.edu/applications/Gator/
\bibitem[\protect\citeauthoryear{De Buizer}{2003}]{d03} De Buizer, J.~M. 2003, \mnras, 341, 277
\bibitem[De Buizer et al. (2000)]{} De Buizer, J.~M., Pi{\~n}a, R.~K., \& Telesco, C.~M.\ 2000, \apjs, 130, 437
\bibitem[De Buizer et al. (2002)]{} De Buizer, J.~M., Walsh, A.~J., Pi{\~n}a, R.~K., Phillips, C.~J., \& Telesco, C.~M.\ 2002, \apj, 564, 327
\bibitem[\protect\citeauthoryear{De Buizer et al.}{2006}]{d06} De Buizer, J.~M. 2006, \apjl, 642, L57
\bibitem[\protect\citeauthoryear{Dodson et al.}{2004}]{d04} Dodson, R., Ojha, R., \& Ellingsen, S.~P.\ 2004, \mnras, 351, 779
\bibitem[\protect\citeauthoryear{Egan et al.}{2003}]{e03} Egan, M.~P., Price, S.~D., Kraemer, K.~E., et al.\ 2003, VizieR Online Data Catalog, 5114, 0
\bibitem[Jayawardhana et al. (1998)]{} Jayawardhana, R., Fisher, R.~S., Hartmann, L., et al.\ 1998, \apjl, 503, L79
\bibitem[\protect\citeauthoryear{Lucy}{1974}]{l74} Lucy, L.~B. 1974, \aj, 79, 745
\bibitem[McCaughrean \& O'Dell (1996)]{} McCaughrean, M.~J., \& O'Dell, C.~R.\ 1996, \aj, 111, 1977
\bibitem[McCaughrean et al. (1998)]{} McCaughrean, M.~J., Chen, H., Bally, J., et al.\ 1998, \apjl, 492, L157
\bibitem[\protect\citeauthoryear{Menten}{1991}]{m91} Menten, K.~M. 1991, \apjl, 380, L75
\bibitem[\protect\citeauthoryear{Minier et al.}{2000}]{m00} Minier, V., Booth, R.~S., \& Conway, J.~E. 2000, \aap, 362, 1093
\bibitem[\protect\citeauthoryear{Norris et al.}{1993}]{n93} Norris, R.~P., Whiteoak, J.~B., Caswell, J.~L., Wieringa, M.~H., Gough, R.~G. 1993, \apj, 412, 222
\bibitem[\protect\citeauthoryear{Norris et al.}{1998}]{n98} Norris, R.~P., Byleveld, S.~E., Diamond, P.~J., et al.\ 1998, \apj, 508, 275
\bibitem[\protect\citeauthoryear{Philips et al.}{1998}]{p98} Philips, C.~J., Norris, R.~P., Ellingsen, S.~P, McCulloch, P.~M. 1998, \mnras, 300, 1131
\bibitem[Reid et al. (2009)]{} Reid, M.~J., Menten, K.~M., Zheng, X.~W., et al.\ 2009, \apj, 700, 137
\bibitem[\protect\citeauthoryear{Richardson}{1972}]{r72} Richardson, W.~H. 1972, J. Opt. Soc. Am., 62, 55
\bibitem[\protect\citeauthoryear{Robitaille et al.}{2007}]{r07}
    Robitaille, T.~P., Whitney, B.~A., Indebetouw, R., \& Wood, K. 2007, \apjs, 169, 328
\bibitem[Rygl et al.(2010)]{} Rygl, K.~L.~J., Wyrowski, F., Schuller, F., \& Menten, K.~M.\ 2010, \aap, 515, A42
\bibitem[Sako et al. (2005)]{} Sako, S., Yamashita, T., Kataza, H., et al.\ 2005, \nat, 434, 995
\bibitem[Schenck et al.(2011)]{} Schenck, D.~E., Shirley, Y.~L., Reiter, M., \& Juneau, S.\ 2011, \aj, 142, 94
\bibitem[Shu \& Adams (1987)]{} Shu, F.~H. \& Adams, F.~C.\ 1987, IAUS, 122, 7
\bibitem[Stecklum \& Kaufl (1998)]{} Stecklum, B. \& Kaufl, H.\ 1998, \textit{ESO Press Release}, PR 08/98
\bibitem[Telesco et al. (1988)]{} Telesco, C.~M., Decher, R., Becklin, E.~E., \& Wolstencroft, R.~D.\ 1988, \nat, 335, 51
\bibitem[\protect\citeauthoryear{Walsh et al.}{1998}]{w98} Walsh, A.~J., Burton, M.~G., Hyland, A.~R., \& Robinson, G.\ 1998, \mnras, 301, 640
\bibitem[Zhang \& Tan (2011)]{} Zhang, Y. \& Tan, J.~C.\ 2011, \apj, 733, 55
\end{thebibliography}
\end{document}